# Revenue Forecasting for Enterprise Products[1]

Amita Gajewar[2], and Gagan Bansal[3].

## Abstract

For any business, planning is a continuous process, and typically business-owners focus on making both long-term planning aligned with a particular strategy as well as short-term planning that accommodates the dynamic market situations. An ability to perform an accurate financial forecast is crucial for effective planning. In this paper, we focus on providing an intelligent and efficient solution that will help in forecasting revenue using machine learning algorithms. We experiment with three different revenue forecasting models, and here we provide detailed insights into the methodology and their relative performance measured on real finance data. As a real-world application of our models, we partner with Microsoft's Finance organization (department that reports Microsoft's finances) to provide them a guidance on the projected revenue for upcoming quarters.

## Introduction

Traditionally, for any company quarterly revenue and/or sales forecast is made by experts in the field who have accumulated domain knowledge in the area of sales and marketing over the years of experience, and using certain statistical methods. These forecasts are built to align with field's modeling for their execution plans. Although, it is crucial to have forecasts made by these experts, we believe that there is a room to make this process more efficient and accurate if we use the machine learning algorithms that can learn from the historical finance data and forecast revenue. Another important advantage of using machine learning algorithms is that it can incorporate social-economic information that is relevant to the business and may correspond to big data sets (several millions of records).

In this paper, we provide insights into the three different machine learning models that we developed using standard time series and regression algorithms. In particular, we use ARIMA (Autoregressive integrated moving average), ETS (Exponential smoothing), STL (Seasonal and trend decomposition using Loess), and Random forest machine learning algorithms to provide the revenue forecast as a guideline to be used by Microsoft Finance in their process of quarterly revenue forecast. Before we describe the algorithms in detail, we mention some of the challenges involved:

- **Limited data set:** For each year, companies will have only 4 quarterly revenue numbers. (Note: it is possible to have daily revenue data logged in, however, we use quarterly data as we want to forecast quarterly revenue).
- **Dynamic market conditions:** The business and economic policy changes in one or many geographical areas may affect the overall sales in these areas, and it is nontrivial to anticipate the impact.

In the following sections we provide formal problem definition, then mention our machine learning models, and then provide the comparative performance of these models.

## Problem Definition:

Given historical quarterly revenue for *m* different geographical areas across the world, we want to forecast the revenue for next quarter (and up to one year in future) for each geographical area as well as the worldwide aggregate (*total*). One may use additional data, e.g., specific products launched in these areas, and potential macro-economic indicators reflecting the market situation (current and past). In order to evaluate the different models we develop, we use the finance data set which corresponds to Microsoft's enterprise products. We perform the segmentation of this data into *m* geographical areas so that we can evaluate the model performance at different regions and worldwide (the entire data set). Considering the confidential nature

---

of the data, we choose not to disclose the details about the logic used to divide the data into different regions and the actual revenue numbers. Further, we provide dummy names for these geographical areas, e.g., *Geo₁*, *Geo₂*, . . . , and *Geoₘ*. Please note that our models can be generalized to data sets where the partitioning is not based on geographical areas. For evaluation purpose though, we partition based on location as it is relevant to most of the multinational companies.

## Revenue Forecasting Models

Now, we describe three different models we developed for revenue forecasting problem.

- **Model 1 -** *Time-Series Model:*

For each geographical area, and worldwide aggregate, with a rolling window-size of *14* quarters, we construct time-series corresponding to revenue-data in those quarters with a seasonality of 4 quarters. Next, we use the functions available in *R* like *auto.arima*, *ets*, *stlf* and *forecast* ([1],[2]), to construct ARIMA, ETS and STL models ([3]) and forecast the revenue for next quarter; let's call these forecasts as *arima-forecast*, ets-*forecast* and *stl-forecast*. Further, we construct derived features using these three forecasts, e.g., we construct *average-ts-forecast* as the mean of *arima-forecast*, ets-*forecast* and *stl-forecast.* Let's assume that the current quarter is q and we want to forecast revenue for quarter (q + 1). In order to choose the best forecast among the time series based forecasts constructed as mentioned previously, we find out which of these models had lowest MAPE (Mean Absolute Percentage Error) for the historical quarters (q, . . ., q-3). We then report forecast obtained from this model as a final forecast for quarter (q + 1). The intuition here is that the model that was able to capture trend and seasonality of the most recent quarters' best is expected to do well in the upcoming quarter as well.

One can extend this model further to include external data, e.g., product launch information in the upcoming and/or recent quarters. One method would be to construct derived features from this external data and the time series based forecast for each quarter, and then generate a linear model with observed revenue as a target. Using the output of linear model and time series based models, one can then construct the final forecast for quarter (q + 1).

- **Model 2 -** *RandomForest Regression Model:*

We develop a more sophisticated model using *randomForest* algorithm in *R* ([4]) that that has richer feature set like: forecasts from standard time-series models (as described in *Model 1*), horizon, geographical-area, and lag-features. Further, using this model we forecast revenue for next four quarters (i.e., four horizons ahead of time). To illustrate this method in detail, let's assume that current quarter is *q* and we want to forecast revenue at quarter (*q + h*) for geographical-area = *A1*, which means that we want to forecast a revenue *h* horizons out from current quarter *q*. To do this, we construct the feature-set as: (1) horizon = *h*, (2) geographical-area = *A1*, (3*)* using historical 16 quarters (q, q-1, …, q-15) of revenue data corresponding to *A1* we build an ARIMA time-series  model and use it to generate the forecast at horizon *h*, let's call it *arima-forecast.* (4) Using past 16 quarters (q, q-1, . . ., q-15) of revenue data corresponding to *A1* we build an ETS time-series model and use it to generate forecast at horizon *h*, let's call it *ets-forecast.* (5) Using past 16 quarters (q, q-1, . . ., q-15) of revenue data corresponding to *A1* we generate forecast at horizon *h* using STL time series model, let's call it *stl-forecast.* (6) Lag-features using historical revenue data up to past eight quarters (q, q-1, . . ., q-8) for geographical-area = *A1*. (7) We also construct derived features using the mean of *arima-forecast*, ets-*forecast* and *stl-forecast* which are computed as mentioned earlier. For most of the experiments we conducted, we set the maximum horizon to four. Thus, with a rolling window-size of 16 quarters, we train the time-series models ARIMA, ETS and STL, and generate forecasts for horizons 1, 2, 3, and 4. Note that, for each geographical area, we train time series models separately and generate forecasts accordingly. This ensures that the fiscal trend particular to a given location is captured. Once we construct a training data corresponding to revenue data for FY (Fiscal Year) 2009-2014 years for all geographical regions at horizons 1 to 4 as described earlier, we train Random Forest model on this entire data set, and that captures the trend worldwide. We attempted to train geo-specific Random Forest models, but found that training on the entire data set yielded better performance in terms of reducing the error in world-wide revenue forecast. Then, using this Random Forest model we obtain forecasts of FY2015 quarterly revenue at horizons 1, 2, 3, and 4. For this we use the *predict* functionality in *R*.

- **Model 3** - *RandomForest Regression Model with Macroeconomic indicators:*

We enhance the feature-set used in *Model 2*, by including the macro-economic indicators: *Total Share Prices for All Shares* and *Current Price Gross Domestic Product* (GDP) for individual geographical areas. We want to incorporate the trend in global market into our model and therefore as a first step we choose to use the GDP and Stock Market data as one of the features into the model. The data for *Total Share Prices for All Shares* is at quarter level, and is adjusted to be in units Index 2010 = 1 whereas the data for *Current Price Gross Domestic Product* is in local currency. For both features, we construct a derived feature by computing year over year growth of quarterly revenue (growth from previous year's same quarter), and therefore there is no need to have standard currency across all regions. Note that, information regarding certain macro-economic indicators like these may not be available for future time-periods, and therefore if we want to use these indicators to construct feature set then we need to first forecast the values of these indicators for the upcoming quarters. Then, we can use these forecasted macro-economic indicators into our revenue forecasting models. In order to achieve this, we use time series based univariate forecasting models (ARIMA) to forecast the data of these macro-economic indicators corresponding to FY2015 and use these forecasts in the test data set as part of the feature-set. As an example, following are the links to macro-economic indicators for the United States: https://research.stlouisfed.org/fred2/series/SPASTT01USQ661N.
https://research.stlouisfed.org/fred2/series/USAGDPNQDSMEI.

## Model Evaluation

- **Experiment 1** – *Evaluation of Model 1 and 2 at horizon 1.*

Using the revenue data for FY2009-2014, we train the models as described by *Models 1* and *2*, and forecast the revenue for FY2015 Q1, Q2, Q3, and Q4 at horizon 1. We then compute the MAPE (Mean Absolute Percentage Error) for FY2015 for individual regions and *total*. Considering the confidential nature of the data we choose not to report actual errors observed, instead we report the performance of *Model 2* relative to *Model 1* in table 1. E.g., if the *Model 1* has an error of x and *Model 2* has an error of y, then the table contains (x-y)/x*100 as an indicator of *Model 2*'s performance compared to *Model 1*.

In table 1, for illustration purpose we report results corresponding to subset of geographical areas. For this we select geographical areas using the following criteria non-exclusively: (a) Top three highest revenue geographical areas ($Geo_1$, $Geo_4$, and $Geo_5$). (b) Top two geographical areas where *Model 1* performs better than Model 2 ($Geo_3$ and $Geo_4$). (c) Top two geographical areas where *Model 2* performs better than *Model 1* ($Geo_1$ and $Geo_2$). (d) *Total* across all geographical areas worldwide.

It can be observed that overall accuracy (*total*) of *Model 2* is better than *Model 1* significantly. Further, we also observed that the *Model 2* performs better for regions with larger revenue. E.g., in the table 1, $Geo_1$ and $Geo_5$ for which Model 2 performs better than *Model 1* constitutes ~44% and ~11% of worldwide revenue for test data set respectively. For regions $Geo_3$ and $Geo_4$, *Model 2* performs significantly worse compared to *Model 1*, however, their total revenue contribution is less than 10% of the worldwide revenue for test data. Therefore, even though the relative performance of *Model 2* is worse for these regions, the worldwide revenue forecasted by *Model 2* is more accurate. Please note that as a first step of the project, our goal is to be more accurate at the worldwide (*total*) level. Regional accuracy is important and improving the accuracy at individual geographical region would be the next step that we would pursue as we continue our work.

**Table 1: Performance of Model 2 relative to Model 1 for FY2015 at horizon 1.**

| Geo-Area | Model 2 |
|---|---|
| $Geo_1$ | 38.30 |
| $Geo_2$ | 43.69 |
| $Geo_3$ | -93.80 |
| $Geo_4$ | -166.67 |
| $Geo_5$ | 11.97 |
| $Geo_6$ | -9.05 |
| Total | 11.45 |

- **Experiment 2:** *Evaluation of Model 2 across multiple horizon*

In experiment 1, it is observed that *Model 2* performs better at minimizing the *total* forecast error at horizon 1. Using this model, we forecast the revenue for FY2015 Q1 to Q4 at horizons 1 to 4, and report the results in table 2. Here, Horizon 1 MAPE corresponds to average APE observed for FY2015 - Q1, Q2, Q3 and Q4 when forecasts are made one quarter out. Horizon 2 MAPE corresponds to average APE observed for FY2015 - Q2, Q3 and Q4 when forecasts are made two quarters out. Similarly, Horizon 3 MAPE corresponds to average APE observed for FY2015 - Q3 and Q4 when forecasts are made three quarters out. And lastly, Horizon 4 MAPE corresponds to average APE for FY2015 - Q4 when forecast is made four quarters out. In the table 2, we report the performance of the model generated using *Model 2* at horizons 2, 3, and 4 compared to its performance at horizon 1. From Table 2, it can be observed that at horizon 2 model performance does not degrade much compared to horizon 1 for most of the individual regions as well as *total*. However, at horizon 4 most of the regions have higher errors compared to horizon 1 forecast. Although, we expect our model to provide more accurate results for immediate quarters as it is crucial for business planning, we would also like to explore methods to reduce the errors observed at longer horizons. Further, please note that for *Geo$_1$* the relative performance of the model at horizon 4 appears to be significantly worse, but this is also because the error observed for *Geo$_1$* at horizon 1 is very small and when it is used as a denominator in the formula to compute relative performance it results into large number. From table 2, it can also be observed that *Model 2* forecasts *total* across all four horizons with reasonable accuracy. Please note that, the *total* forecast error observed at every horizon from one to four for the test data set FY2015, is less than 3.5% (due to confidentiality exact numbers cannot be shared). This is a good indicator that *Model 2* generates forecasts that are reliable for longer horizons for worldwide aggregate.

**Table 2: Performance of Model 2 for FY2015 at horizons 2, 3 and 4 relative to horizon 1.**

| Geo-Area | Horizon 2 | Horizon 3 | Horizon 4 |
|---|---|---|---|
| *Geo$_1$* | -37.93 | 56.16 | -322.66 |
| *Geo$_2$* | 14.05 | -26.48 | 21.18 |
| *Geo$_3$* | -4.53 | -48.67 | -72.02 |
| *Geo$_4$* | 3.88 | -30.00 | -81.63 |
| *Geo$_5$* | 3.47 | -16.00 | -44.27 |
| *Geo$_6$* | -13.02 | 32.99 | -36.28 |
| Total | 53.63 | 52.68 | 76.66 |

- **Experiment 3 –** *Evaluation of Model 3 across multiple horizons.*

As described earlier, in *Model 3* we include additional features corresponding to the macro-economic indicators. We develop two models, *Model3_stock* and *Model3_gdp* using the macro-economic indicators *Total Share Prices for All Shares* and *Current Price GDP*, respectively. In order to evaluate the model performance, similar to experiment 2, we use *Model3_stock* and *Model3_gdp* to forecast the revenue for FY2015 Q1 to Q4 at horizons 1 to 4.

**Table 3A: Performance of Model 3 for FY2015 relative to Model 2.**

| Geo-Area = Geo1 | Horizon 1 | Horizon 2 | Horizon 3 | Horizon 4 |
|---|---|---|---|---|
| Model3_stock | -51.72 | -35.36 | -62.92 | -12.94 |
| Model3_gdp | -30.05 | -20.36 | -26.97 | 1.05 |

**Table 3B: Performance of Model 3 for FY2015 relative to Model 2.**

| Geo-Area= Total | Horizon 1 | Horizon 2 | Horizon 3 | Horizon 4 |
|---|---|---|---|---|
| Model3_stock | -9.15 | -31.29 | -6.00 | -141.89 |
| Model3_gdp | 5.68 | 8.16 | -50.67 | -51.35 |

We run the experiments for all regions, but here we report the results for *Geo$_1$* (largest in terms of revenue) and *total* to give an idea of the model's performance. The results are reported in tables 3A and 3B, using the *Model 2*'s performance at horizons 1 to 4 as a baseline for respective horizons. We observed similar behavior for other regions as well. It can be inferred from the tables 3A and 3B that the inclusion of this data does not necessarily contribute towards improving the accuracy of the model developed using *Model 2*. We also observed weak correlation between these macro-economic indicators and quarterly revenue observed for the FY2009-2015 data set that we used for experiments. This weak correlation also explains the lack of significant improvement in *Model 3*'s performance compared to *Model 2*. Although results are not promising for the macro-economic indicators we tried, the methodology developed to use these indicators can be easily extended to include any other external data. As part of future work, we plan on using more relevant external information to construct features for *Model 3*.

## Real World Application

We have been using our models to provide the guidance to Microsoft's major finance divisions for the past three quarters. For model evaluation purpose, we ran all the experiments using the revenue data corresponding to enterprise products, however, our model is generic to be used for other finance divisions with similar business model. The finance divisions for which we have provided forecasts corresponds to more than 50% of Microsoft's quarterly revenue altogether. In the table 4, we provide the comparison of the model's performance with the expert-judgements corresponding to Microsoft's enterprise products finance division data set. Although, we provide forecasts for several geographical regions, here in table 4, we provide results corresponding to forecasts of worldwide aggregate revenue, computed at horizon 1. E.g., we use the revenue data until FY2016-Q3 to train the model and then FY2016-Q4 quarterly revenue forecast is computed at horizon 1. As mentioned previously, we report the improvement of machine learning model compared to expert-judgement forecast using the formula ($APE_{exp}$ − $APE_{ml}$)/$APE_{exp}$ * 100 where $APE_{exp}$ and $APE_{ml}$ are the absolute percentage errors corresponding to expert-judgement forecast and machine learning model forecast, respectively. As seen in table 4, the machine learning model has outperformed expert-judgement forecasts significantly for all three quarters in a row. Please note that all expert-judgement and machine learning model APEs are less than 5% for FY2016-Q2, Q3 and Q4. Since machine learning models show an improvement over expert judgement forecasts that are very accurate, it further gives credence to our methodology and a promising solution machine learning has to offer in the area of finance forecasting.

**Table 4: Machine learning model improvement relative to human judgement.**

| Geo-Area= Total | Horizon 1 |
|---|---|
| FY16-Q2 | 53% |
| FY16-Q3 | 10% |
| FY16-Q4 | 90% |

## Conclusion

Initial experiments suggest that the models developed using machine learning algorithms can forecast quarterly revenue with reasonable accuracy. The improvement in *total* forecast error for *Model 2* compared to *Model 1* indicates a progress in right direction. Further, our methodology can incorporate social-economic information and key business drivers easily, and scale well with relevant big data. The quarterly forecasts computed using our models have been adopted by Microsoft's finance team including the CFO (Chief Finance Officer). For some of the revenue forecasts we computed, we have observed an error as low as 0.1%.

As a future work, we would like to experiment with adding more relevant auxiliary data (e.g., sales plans) that would provide better indicators for revenue. We also plan to explore methods to minimize the errors for various geographical regions by taking into account geo-specific features while building the models. We believe that using machine learning models for revenue forecasting that provide reasonable accuracy will be very helpful for finance organizations as it will be free of human judgements and computed in an efficient manner. In future, we would also like to provide an end-to-end automated forecasting solution in Azure.

## Acknowledgement
We would like to thank Jocelyn Barker, Matt Conners, Konstantin Golyaev, Mike Keller, Natasha Milovic, and Meng-Hua Wei for many helpful discussions.